\documentclass[fleqn]{article}
\usepackage{amscd,amsmath,amssymb,emlines2}
\usepackage[dvips]{graphicx}
\title{Isometry classification of cubic homogeneous 3-dimensional forms}
\author{Sergey S. Kokarev\thanks{logos-center@mail.ru}}
\date{Research Institute of Hypercomplex System in Geometry and Physics (Moscow),\\
RSEC "Logos"\,(Yaroslavl), Russia}

\begin{document}

\maketitle

\begin{abstract}
The problem of classification of cubic homogeneous Finslerian
3D metrics with respect to their isometries is considered.
It is shown, that there are 6 different general affine types of such metrics.
Algebras of isometries are presented in apparent kind together with their
affine-invariant properties. Interrelation between symmetries and
projective classifyings is discussed.
\end{abstract}

\section{Introduction}

One of the main tools for studying invariant  geometrical
properties of manifolds is symmetry considered in a wide
sense of the word. The most important and simple kind of
symmetry is isometry, which can be defined for any manifold with metric $G$.
Local definition of continious (or, more correctly, smooth) isometry
involves vector field $X$ ({\it Killing field}), satisfying the
following {\it Killing equation}:
\begin{equation}\label{kill}
L_XG=0,
\end{equation}
where $L_X$ denotes standard Lie derivative along $X$ \cite{warn}.
It is well known, that isometries fields form Lie algebra with
respect to Lie bracket, and invariant properties of the Lie algebra
(dimension, solvability, presence or absence of subalgebras and ideals etc.)
express invariant properties of the manifolds itself. Often
Lie algebra of isometries defines metrics $G$ uniquely or up to some
arbitrary functions. For example, homogeneous space-times in cosmological models of
GR admits complete Bianchi classification, which is convenient mean for classifying both of the models itself and
their important physical properties \cite{shm,land}.

Isometries of manifolds with homogeneous quadratic metrics (i.e. satisfying
$G_{\alpha\beta}=\text{const}$), are well known. If we exclude
degenerate cases, we always deals with metrics of: a) euclidean
type; b) pseudoeuclidean type; c) symplectic type d) mixed type,
containing all or part of the types (a), (b), (c) in the form of
their direct sum. In the case (a) we have isometry groups $O(n),$ in the case
(b) ---  groups $O(m,n),$ in the case (c)  --- groups $\text{Sp}(n)$
for some natural $n,m.$ In case (d) we have some combinations of
above listed groups. These types of metrics together with their groups of isometries form local geometric
base for construction of more complicated physical models of
space-time, relevant to some modern physical concepts (say, gauge principle) and
experimental data.

Last decade we observe increasing interest both from physics and from mathematics  for geometrical models
involving nonquadratic metrics of Finslerian kind \cite{bogosl,pavl,sip}. Such models are of
great interest, since they naturally reflects possible anisotropy of
space-time. The hypothesis about anisotropy of space-time  makes more clear some experimental data in
cosmology, astrophysics and elementary particles physics.

In view of further development of Finslerian geometrical models we
need more deeper understanding of their geometrical properties.
Present paper is devoted to investigation of what we can know about
homogeneous cubic metrics by methods of Lie isometries theory. We
restrict ourselves by cubic metrics in 3D space, which play important role for hypercomplex numbers\footnote{Present paper is developed version of
the part of lectures, delivered by author on autumn-2008 School on Finslerian Geometry and Hypercomplex Numbers (Fryazino, Russia). Full text of the lectures
will be published in 2009 in special issue of Research Institute of Hypercomplex Systems in Geometry and Physics (www.polynumbers.ru).}.
Our analysis shows, that in a difference with isometries of quadratic
homogeneous metrics in 3D 1) isometries of homogeneous cubic metrics form
more rich family (for nondegenerate metrics 5 cubic against 3
quadratic); 2) isometry classification is not complete, since some
different projective classes of cubic metrics belong to the same
symmetry class (see table at the end of the section \ref{projj}).

These results (obtained for the simplest class of Finslerian metrics)
at least show, that classical Lie analysis is useful but unsufficient for relevant understanding of Finslerian models and more subtle aspects of symmetry should be
incorporated in this topic.

\section{Metrics and affine types}

We are going to investigate isometries of homogenious cubic
metrics of the form:
\begin{equation}\label{cub1}
G=G_{\alpha\beta\gamma}dx^\alpha\otimes dx^\beta\otimes dx^\gamma,
\end{equation}
where $G_{\alpha\beta\gamma}$ --- symmetric real cubic matrix.
Geometrical spaces with metric of such type are commonly referred to
{\it Finslerian spaces} and metrics $G$ is commonly related to special  {\it Finslerian metric} \cite{rund}.
Let us introduce the following notations:
\begin{equation}\label{design}
G_{\alpha\alpha\alpha}= A_\alpha;\quad G_{122}=B_1;\quad  G_{133}=B_2;\quad
G_{233}=B_3;\quad
\end{equation}
\[
G_{112}=C_1;\quad G_{113}=C_2;\quad G_{223}=C_3;\quad G_{123}=F,
\]
where all $A_\alpha,$ $B_\beta,$ $C_\gamma$ and $F$ are constants.

Not all of the components (\ref{design}) have geometric significance.
Representation (\ref{design}) is invariant with respect to choice of coordinate systems
within class  of affine-equivalent ones, where all components of $G$
are constant.
Any matrix of nondegenerate affine homogeneous transformation in
$R^3$ has in general 9 independent components, which could be used so, that
9 from 10 components of $G$ will vanish.

So, we can preliminary conclude, that:
\begin{enumerate}
\item
For complete investigation of the problem it is sufficient to
consider metrics with some small number of nonzero components;
\item
It is necessary to investigate all possible combinations of these
components.
\end{enumerate}
We will show, that it is, in fact, sufficiently to study the metrics $G$ with nonzero components of number no greater then 6.

The number of nonzero coefficients of homogeneous metric $G$ will define
{\it its affine type  $\tau(G)$}. Note, that affine type of metric  $G$ depends on choice of affine coordinate system.
Invariant characteristic, independent on choice of affine coordinates,
is  {\it exact affine type:}
\[
\tau_0(G)\equiv\min\limits_{\text{Aff}(R^3)}\tau(G),
\]
where $\text{Aff}(R^3)$ --- class of affine coordinate system in $R^3,$
connected by nondegenerate affine transformations.
Lets call two homogeneous metrics  $G_1$ and $G_2$ {\it equivalent}: $G_1\sim G_2,$
if there exist such homogeneous nondegenerate affine transformation in $R^3$,  which transforms
$G_1$ into $G_2$ or vice verse. Obviously,  for equivalent metrics:
$G_1\sim G_2$ may be $\tau(G_1)\neq\tau(G_2),$ but it is necessarily must be:
$\tau_0(G_1)=\tau_0(G_2).$ However, coincidence of exact affine types for some two
metrics, generally speaking, is not sufficient for their equivalence, since
the components, which compose minimal sets of nonzero ones  may be different for these
two metrics.

\section{Killing equations and their solutions}

General system of Killing equations  (\ref{kill}) takes the form:
\begin{equation}\label{killG}
\left\{
\begin{array}{l}
3A_1\partial_1X^1+3C_1\partial_1X^2+3C_2\partial_1X^3 = 0;\\
2C_1\partial_1X^1+2B_1\partial_1X^2+2F\partial_1X^3+A_1\partial_2X^1+C_1\partial_2X_2+
C_2\partial_2X^3=0;\\
B_1\partial_1X^1+A_2\partial_1X^2+C_3\partial_1X^3+2C_1\partial_2X^1+2B_1\partial_2X^2
+2F\partial_2X^3=0;\\
3B_1\partial_2X^1+3A_2\partial_2X^2+3C_3\partial_2X^3=0;\\
2C_2\partial_1X^1+2F\partial_1X^2+2B_2\partial_1X^3+A_1\partial_3X^1+C_1\partial_3X^2+C_2\partial_3X^3
 = 0;\\
F\partial_1X^1+C_3\partial_1X^2+B_3\partial_1X^3+C_2\partial_2X^1+F\partial_2X^2+B_2\partial_2
X^3+C_1\partial_3X^1+\\ +B_1\partial_3X^2+F\partial_3X^3=0;\\
2F\partial_2X^1+2C_3\partial_2X^2+2B_3\partial_2X^3+B_1\partial_3X^1+A_2\partial_3X^2+C_3\partial_3X^3= 0;\\
B_2\partial_1X^1+B_3\partial_1X^2+A_3\partial_1X^3+2C_2\partial_3X^1+2F\partial_3X^2+
2B_2\partial_3X^3=0;\\
B_2\partial_2X^1+B_3\partial_2X^2+A_3\partial_2X^3+2F\partial_3X^1+2C_3\partial_3X^2+2B_3\partial_3X^3=0;\\
3B_2\partial_3X^1+3B_3\partial_3X^2+3A_3\partial_3X^3=0.
\end{array}
\right.
\end{equation}

Let us consider consequently all cases of general metrics with
different $\tau(G).$
Everywhere we'll use freedom of scales of coordinates for
transforming of maximal number of components
into $\pm1$ (canonical kind). We'll not consider separately those of the cases, which differ from each other
by permutations of coordinates.
Also we'll  omit constant vector fields of isometries, forming subalgebra of translations of complete algebra of isometries of $G$,
and will focus only on symmetries, different from translations.
We shall call them  {\it nontrivial symmetries} of homogeneous Finslerian metrics.

\subsection{Metrics with $\tau(G)=\tau_0(G)=1 $ (3 types)}

In notation  of different cases only nonzero components of metric are
shown (all remaining are zero).
Hereafter we list only the cases with nontrivial symmetries.
\begin{enumerate}
\item $F\neq0.$ Canonical form of metric:
\begin{equation}\label{bm3}
G=\hat{\mathcal{S}}(dx^1\otimes dx^2\otimes dx^3),
\end{equation}
where $\hat{\mathcal{S}}$ --- tensor product symmetrization operator.
This metric is known as  {\it Berwald-Moor metric.}
Nontrivial symmetries are \cite{gar,kok}:
\[
X_1=x^1\partial_1-x^2\partial_2;\quad
X_2=x^1\partial_1-x^3\partial_3.
\]
These are unimodular dilatations of coordinate axes.
Note, that this algebra is abelian.
\item $B_1\neq0.$ Canonical form of metric:
\begin{equation}\label{deg2}
G=dx^1\otimes dx^2\otimes dx^2+dx^2\otimes dx^1\otimes dx^2+dx^2\otimes dx^2\otimes dx^1.
\end{equation}
Algebra of nontrivial symmetries is infinitely dimensional:
\[
X=x^2\partial_2-2x^1\partial_1+f(x^1,x^2,x^3)\partial_3,
\]
where $f$ --- arbitrary smooth function of three variables.
\item $A_1\neq0.$ Canonical form of metric:
\begin{equation}\label{degg2}
G=dx^1\otimes dx^1\otimes dx^1.
\end{equation}
Algebra of nontrivial symmetries is infinitely dimensional:
\[
X=f_2(x^1,x^2,x^3)\partial_2+f_3(x^1,x^2,x^3)\partial_3,
\]
where $f_2,f_3$ --- arbitrary smooth functions of three variables.
\end{enumerate}

These three cases exaust nontrivial cases of the class $\tau_0(G)=1.$
Note, that metrics (\ref{deg2})-(\ref{degg2}) are degenerated,
since they are described by subspaces of 3-dimensional basis of 1-forms $\{dx^1,dx^2,dx^3\}.$

\subsection{Metrics with $\tau(G)=2$ (9 types)}

\begin{enumerate}
\item $F\neq0,A_1\neq0.$ Canonical form of metric:
\begin{equation}\label{bm3s1}
G=dx^1\otimes dx^1\otimes dx^1+\hat{\mathcal{S}}(dx^1\otimes dx^2\otimes dx^3).
\end{equation}
Algebra of nontrivial symmetries is 1-dimensional:
\[
X=x^2\partial_2-x^3\partial_3.
\]
\item $F\neq0,B_1\neq0.$ Canonical form of metric:
\begin{equation}\label{bm3s2}
G=dx^1\otimes dx^2\otimes dx^2+dx^2\otimes dx^1\otimes dx^2+dx^2\otimes dx^2\otimes dx^1
+\hat{\mathcal{S}}(dx^1\otimes dx^2\otimes dx^3).
\end{equation}
Algebra of nontrivial symmetries is 2-dimensional:
\[
X_1=x^1\partial_1-(x^3+x^2/2)\partial_3;\quad x^2\partial_2-(x^3+x^2)\partial_3.
\]
\item $A_1\neq0,B_3\neq0.$ Canonical form of metric:
\begin{equation}\label{bms3}
G=dx^1\otimes dx^1\otimes dx^1+dx^2\otimes dx^3\otimes dx^3+dx^3\otimes dx^2\otimes dx^3+
dx^3\otimes dx^3\otimes dx^2.
\end{equation}
Algebra of nontrivial symmetries is 1-dimensional:
\[
X=x^2\partial_2-(x^3/2)\partial_3.
\]
\item $A_1\neq0,C_1\neq0.$ Canonical form of metric:
\begin{equation}\label{bms4}
G=dx^1\otimes dx^1\otimes dx^1+dx^1\otimes dx^1\otimes dx^3+dx^1\otimes dx^3\otimes dx^1+
dx^3\otimes dx^1\otimes dx^1.
\end{equation}
Algebra of nontrivial symmetries is 1-dimensional:
\[
X=x^1\partial_1-(2x^2+x^1)\partial_2.
\]
\item $B_1\neq0,B_2\neq0.$ Canonical form of metric:
\begin{equation}\label{bms5}
G=dx^1\otimes dx^2\otimes dx^2+dx^2\otimes dx^1\otimes dx^2+dx^2\otimes dx^2\otimes dx^1
\end{equation}
\[
\pm
(dx^1\otimes dx^3\otimes dx^3+dx^3\otimes dx^1\otimes dx^3+dx^3\otimes dx^3\otimes dx^1).
\]
Algebra of nontrivial symmetries is 1-dimensional:
\[
X=x^1\partial_1-(x^2/2)\partial_2-(x^3/2)\partial_3.
\]
\item $B_1\neq0,B_3\neq0.$ Canonical form of metric:
\begin{equation}\label{bms6}
G=dx^1\otimes dx^2\otimes dx^2+dx^2\otimes dx^1\otimes dx^2+dx^2\otimes dx^2\otimes dx^1+
\end{equation}
\[
dx^2\otimes dx^3\otimes dx^3+dx^3\otimes dx^3\otimes dx^2+dx^3\otimes dx^2\otimes dx^3.
\]
Algebra of nontrivial symmetries is 1-dimensional:
\[
X_1=-2x^1\partial+x^2\partial_2-(x^3/2)\partial_3;\quad
X_2=x^3\partial_1-(x^2/2)\partial_3.
\]
\item $B_1\neq0,C_3\neq0.$ Canonical form of metric:
\begin{equation}\label{bms7}
G=dx^1\otimes dx^2\otimes dx^2+dx^2\otimes dx^1\otimes dx^2+dx^2\otimes dx^2\otimes dx^1+
\end{equation}
\[
dx^3\otimes dx^2\otimes dx^2+dx^2\otimes dx^3\otimes dx^2+dx^2\otimes dx^2\otimes dx^3.
\]
Algebra of nontrivial symmetries is $\infty$-dimensional:
\[
X_1=x^2\partial_2-2(x^1+x^3)\partial_3;\quad
X_2=f(x^1,x^2,x^3)(\partial_1-\partial_3),
\]
where $f$ --- arbitrary smooth function of three variables.
\item $A_1\neq0,A_2\neq0.$ Canonical form of metric:
\begin{equation}\label{bmss7}
G=dx^1\otimes dx^1\otimes dx^1+dx^2\otimes dx^2\otimes dx^2
\end{equation}
Algebra of nontrivial symmetries is $\infty$-dimensional:
\[
X=f(x^1,x^2,x^3)\partial_3,
\]
where $f$ --- arbitrary smooth function of three variables.
\item $A_1\neq0,B_1\neq0.$ Canonical form of metric:
\begin{equation}\label{bmsss7}
G=dx^1\otimes dx^1\otimes dx^1\pm\hat{\mathcal{S}}(dx^1\otimes dx^2\otimes
dx^2)
\end{equation}
Algebra of nontrivial symmetries is $\infty$-dimensional:
\[
X=f(x^1,x^2,x^3)\partial_3,
\]
where $f$ --- arbitrary smooth function of three variables.
\end{enumerate}

\subsection{Metrics with $\tau(G)=3$ (13 types)}

In majority of the cases symmetries are trivial.
Only 13 metrics possess nontrivial symmetries.

\begin{enumerate}
\item $F\neq0,A_1\neq0,B_1\neq0.$ Canonical form of metric:
\begin{equation}\label{bm3s8}
G=dx^1\otimes dx^1\otimes dx^1\pm\hat{\mathcal{S}}(dx^1\otimes dx^2\otimes dx^2)+\hat{\mathcal{S}}(dx^1\otimes dx^2\otimes dx^3).
\end{equation}
Algebra of nontrivial symmetries is 1-dimensional:
\[
X=x^2\partial_2-(x^3\pm x^2)\partial_3.
\]
\item $F\neq0,A_1\neq0,C_1\neq0.$ Canonical form of metric:
\begin{equation}\label{bm3s9}
G=dx^1\otimes dx^1\otimes dx^1+\hat{\mathcal{S}}(dx^1\otimes dx^1\otimes dx^2)+\hat{\mathcal{S}}(dx^1\otimes dx^2\otimes dx^3).
\end{equation}
Algebra of nontrivial symmetries is 1-dimensional:
\[
X=x^2\partial_2-(x^3+x^1/2)\partial_3.
\]
\item $F\neq0,B_1\neq0,B_2\neq0.$ Canonical form of metric:
\begin{equation}\label{bms10}
G=\hat{\mathcal{S}}(dx^1\otimes dx^2\otimes dx^2)\pm\hat{\mathcal{S}}(dx^1\otimes dx^3\otimes dx^3)+\hat{\mathcal{S}}(dx^1\otimes dx^2\otimes dx^3).
\end{equation}
Algebra of nontrivial symmetries is 2-dimensional:
\[
X_1=x^1\partial_1\pm(x^3/2)\partial_2-(x^3+x^2/2)\partial_3;\quad
X_2=(x^2\pm x^3)\partial_2-(x^2+x^3)\partial_3.
\]
\item $F\neq0,B_1\neq0,B_3\neq0.$ Canonical form of metric:
\begin{equation}\label{bms11}
G=\hat{\mathcal{S}}(dx^1\otimes dx^2\otimes dx^2)+\hat{\mathcal{S}}(dx^2\otimes dx^3\otimes dx^3)+\hat{\mathcal{S}}(dx^1\otimes dx^2\otimes dx^3).
\end{equation}
Algebra of nontrivial symmetries is 1-dimensional:
\[
X=(x^1+x^3)\partial_1-(x^3+x^2/2)\partial_3.
\]
\item $F\neq0,B_1\neq0,C_1\neq0.$ Canonical form of metric:
\begin{equation}\label{bms12}
G=\hat{\mathcal{S}}(dx^1\otimes dx^2\otimes dx^2)+\hat{\mathcal{S}}(dx^1\otimes dx^1\otimes dx^2)+\hat{\mathcal{S}}(dx^1\otimes dx^2\otimes dx^3).
\end{equation}
Algebra of nontrivial symmetries is 2-dimensional:
\[
X_1=x^2\partial_2-(x^2+x^3)\partial_3;\quad
X_2=x^1\partial_1-(x^3+x^2/2+x^1)\partial_3.
\]
\item $F\neq0,B_1\neq0,C_3\neq0.$ Canonical form of metric:
\begin{equation}\label{bms13}
G=\hat{\mathcal{S}}(dx^1\otimes dx^2\otimes dx^2)+\hat{\mathcal{S}}(dx^2\otimes dx^2\otimes dx^3)+\hat{\mathcal{S}}(dx^1\otimes dx^2\otimes dx^3).
\end{equation}
Algebra of nontrivial symmetries is 1-dimensional:
\[
X=(x^3+x^2/2)\partial_3-(x^1+x^2/2)\partial_1.
\]
\item $A_1\neq0,A_2\neq0,C_2\neq0.$ Canonical form of metric:
\begin{equation}\label{bms14}
G=dx^1\otimes dx^1\otimes dx^1+dx^2\otimes dx^2\otimes dx^2+\hat{\mathcal{S}}(dx^1\otimes dx^1\otimes dx^3).
\end{equation}
Algebra of nontrivial symmetries is 1-dimensional:
\[
X=x^1\partial_1-(x^1+2x^3)\partial_3.
\]
\item $A_1\neq0,B_1\neq0,B_2\neq0.$ Canonical form of metric:
\begin{equation}\label{bms15}
G=dx^1\otimes dx^1\otimes dx^1+\epsilon_1\hat{\mathcal{S}}(dx^1\otimes dx^2\otimes dx^2)+\epsilon_2\hat{\mathcal{S}}(dx^1\otimes dx^3\otimes
dx^3),
\end{equation}
where $\epsilon_1=\pm1,$ $\epsilon_2=\pm1$ ---  independent sign factors.
Algebra of nontrivial symmetries is 1-dimensional:
\[
X=x^3\partial_2-\epsilon_1\epsilon_2x^2\partial_3.
\]
\item $A_1\neq0,B_1\neq0,C_2\neq0.$ Canonical form of metric:
\begin{equation}\label{bms16}
G=dx^1\otimes dx^1\otimes dx^1\pm\hat{\mathcal{S}}(dx^1\otimes dx^2\otimes dx^2)+\hat{\mathcal{S}}(dx^1\otimes dx^1\otimes
dx^3),
\end{equation}
Algebra of nontrivial symmetries is 2-dimensional:
\[
X_1=x^1\partial_1-(x^2/2)\partial_2-2x^3\partial_3;\quad
X_2=\mp(x^1/2)\partial_2+x^2\partial_3.
\]
\item $A_1\neq0,B_1\neq0,C_3\neq0.$ Canonical form of metric:
\begin{equation}\label{bms17}
G=dx^1\otimes dx^1\otimes dx^1\pm\hat{\mathcal{S}}(dx^1\otimes dx^2\otimes dx^2)+\hat{\mathcal{S}}(dx^2\otimes dx^2\otimes
dx^3),
\end{equation}
Algebra of nontrivial symmetries is 1-dimensional:
\[
X=x^2\partial_2-2(x^3\pm x^1)\partial_3.
\]
\item $B_1\neq0,B_2\neq0,C_1\neq0.$ Canonical form of metric:
\begin{equation}\label{bms18}
G=\hat{\mathcal{S}}(dx^1\otimes dx^2\otimes dx^2)\pm\hat{\mathcal{S}}(dx^1\otimes dx^3\otimes dx^3)+\hat{\mathcal{S}}(dx^1\otimes dx^1\otimes
dx^2),
\end{equation}
Algebra of nontrivial symmetries is 1-dimensional:
\[
X=x^3\partial_2\mp 2(x^1/2+ x^2)\partial_3.
\]
\item $B_1\neq0,B_3\neq0,C_1\neq0.$ Canonical form of metric:
\begin{equation}\label{bms19}
G=\hat{\mathcal{S}}(dx^1\otimes dx^2\otimes dx^2)+\hat{\mathcal{S}}(dx^2\otimes dx^3\otimes dx^3)+\hat{\mathcal{S}}(dx^1\otimes dx^1\otimes
dx^2).
\end{equation}
Algebra of nontrivial symmetries is 1-dimensional:
\[
X=x^3\partial_1-(x^1+ x^2/2)\partial_3.
\]
\item $B_1\neq0,B_3\neq0,C_3\neq0.$ Canonical form of metric:
\begin{equation}\label{bms20}
G=\hat{\mathcal{S}}(dx^1\otimes dx^2\otimes dx^2)\pm\hat{\mathcal{S}}(dx^2\otimes dx^3\otimes dx^3)+\hat{\mathcal{S}}(dx^2\otimes dx^2\otimes
dx^3).
\end{equation}
Algebra of nontrivial symmetries is 2-dimensional:
\[
X_1=x^2\partial_2-(x^3/2)\partial_3-(2x^1+3x^3/2)\partial_1;\quad
X_2=x^2\partial_3-(x^2\pm 2x^3)\partial_1.
\]
\end{enumerate}

\subsection{Metrics with $\tau(G)=4$ (10 types)}

\begin{enumerate}
\item $F\neq0,A_1\neq0,B_1\neq0,B_2\neq0.$ Canonical form of metric:
\begin{equation}\label{bm3s21}
G=F\hat{\mathcal{S}}(dx^1\otimes dx^2\otimes dx^3)+\epsilon_1\hat{\mathcal{S}}(dx^1\otimes dx^2\otimes dx^2)+\epsilon_2\hat{\mathcal{S}}(dx^1\otimes dx^3\otimes dx^3)
+dx^1\otimes dx^1\otimes dx^1.
\end{equation}
Algebra of nontrivial symmetries is 1-dimensional:
\[
X=(x^3+\epsilon_2Fx^2)\partial_2-\epsilon_2(\epsilon_1x^2+Fx^3)\partial_3.
\]
\item $F\neq0,A_1\neq0,B_1\neq0,C_2\neq0.$ Canonical form of metric:
\begin{equation}\label{bm3s22}
G=F\hat{\mathcal{S}}(dx^1\otimes dx^2\otimes dx^3)\pm\hat{\mathcal{S}}(dx^1\otimes dx^2\otimes dx^2)+\hat{\mathcal{S}}(dx^1\otimes dx^1\otimes dx^3)
+dx^1\otimes dx^1\otimes dx^1.
\end{equation}
Algebra of nontrivial symmetries is 1-dimensional:
\[
X=(x^3\pm x^2/F)\partial_3-(x^2+x^1/2F)\partial_2.
\]
\item $F\neq0,B_1\neq0,B_2\neq0,C_2\neq0.$ Canonical form of metric:
\begin{equation}\label{bm3s23}
G=F\hat{\mathcal{S}}(dx^1\otimes dx^2\otimes dx^3)\pm\hat{\mathcal{S}}(dx^1\otimes dx^2\otimes dx^2)+\hat{\mathcal{S}}(dx^1\otimes dx^3\otimes dx^3)
+\hat{\mathcal{S}}(dx^1\otimes dx^1\otimes dx^3).
\end{equation}
Algebra of nontrivial symmetries is 1-dimensional:
\[
X=(x^2\pm Fx^3)\partial_3\mp(Fx^2+x^3+x^1/2)\partial_2.
\]
\item $F\neq0,B_2\neq0,B_3\neq,C_2\neq0.$ Canonical form of metric:
\begin{equation}\label{bm3s24}
G=F\hat{\mathcal{S}}(dx^1\otimes dx^2\otimes dx^3)+\hat{\mathcal{S}}(dx^1\otimes dx^3\otimes dx^3)+\hat{\mathcal{S}}(dx^2\otimes dx^3\otimes dx^3)
+\hat{\mathcal{S}}(dx^1\otimes dx^1\otimes dx^3).
\end{equation}
Algebra of nontrivial symmetries is 1-dimensional:
\[
X=(x^1+x^3/2F)\partial_1-(x^2+x^1/F+x^3/2F)\partial_2.
\]
\item $A_1\neq0,A_2\neq0,B_1\neq0,C_2\neq0.$ Canonical form of metric:
\begin{equation}\label{bm3s25}
G=dx^1\otimes dx^1\otimes dx^1+dx^2\otimes dx^2\otimes dx^2+B\hat{\mathcal{S}}(dx^1\otimes dx^2\otimes dx^2)
+\hat{\mathcal{S}}(dx^1\otimes dx^1\otimes dx^3).
\end{equation}
Algebra of nontrivial symmetries is 1-dimensional:
\[
X=x^1\partial_1-Bx^1\partial_2+2(B^2x^2-x^3-x^1/2)\partial_3.
\]
\item $A_1\neq0,A_2\neq0,B_1\neq0,C_3\neq0.$ Canonical form of metric:
\begin{equation}\label{bm3s26}
G=dx^1\otimes dx^1\otimes dx^1+dx^2\otimes dx^2\otimes dx^2+B\hat{\mathcal{S}}(dx^1\otimes dx^2\otimes dx^2)
+\hat{\mathcal{S}}(dx^2\otimes dx^2\otimes dx^3).
\end{equation}
Algebra of nontrivial symmetries is 1-dimensional:
\[
X=x^2\partial_2-(2x^3+2Bx^1+x^2)\partial_3.
\]
\item $A_1\neq0,B_1\neq0,B_2\neq0,C_1\neq0.$ Canonical form of metric:
\begin{equation}\label{bm3s27}
G=dx^1\otimes dx^1\otimes dx^1+\epsilon_1\hat{\mathcal{S}}(dx^1\otimes dx^2\otimes dx^2)+
\epsilon_2\hat{\mathcal{S}}(dx^1\otimes dx^3\otimes dx^3)
+C\hat{\mathcal{S}}(dx^1\otimes dx^1\otimes dx^3).
\end{equation}
Algebra of nontrivial symmetries is 1-dimensional:
\[
X=x^3\partial_2-(\epsilon_1\epsilon_2x^2+\epsilon_2Cx^1/2)\partial_3.
\]
\item $A_1\neq0,B_2\neq0,B_3\neq0,C_2\neq0.$ Canonical form of metric:
\begin{equation}\label{bm3s28}
G=dx^1\otimes dx^1\otimes dx^1\pm\hat{\mathcal{S}}(dx^1\otimes dx^3\otimes dx^3)+
\hat{\mathcal{S}}(dx^2\otimes dx^3\otimes dx^3)
+C\hat{\mathcal{S}}(dx^1\otimes dx^1\otimes dx^3).
\end{equation}
Algebra of nontrivial symmetries is 1-dimensional:
\[
X=x^3\partial_3-Cx^3\partial_1+(\pm Cx^3+2(C^2\mp1)x^1-2x^2)\partial_2.
\]
\item $A_1\neq0,B_1\neq0,C_1\neq0,C_2\neq0.$ Canonical form of metric:
\begin{equation}\label{bm3s29}
G=dx^1\otimes dx^1\otimes dx^1+B\hat{\mathcal{S}}(dx^1\otimes dx^2\otimes dx^2)+
\hat{\mathcal{S}}(dx^1\otimes dx^1\otimes dx^2)
+\hat{\mathcal{S}}(dx^1\otimes dx^1\otimes dx^3).
\end{equation}
Algebra of nontrivial symmetries is 2-dimensional:
\[
X_1=x^1\partial_1-(x^2/2)\partial_2-(2x^3+x^1+3x^2/2)\partial_3;\quad
X_2=x^1\partial_2-(x^1+2Bx^2)\partial_3.
\]
\item $B_1\neq0,B_2\neq0,C_1\neq0,C_2\neq0.$ Canonical form of metric:
\begin{equation}\label{bm3s30}
G=B\hat{\mathcal{S}}(dx^1\otimes dx^2\otimes dx^2)+\hat{\mathcal{S}}(dx^1\otimes dx^3\otimes dx^3)+
\hat{\mathcal{S}}(dx^1\otimes dx^1\otimes dx^2)
+\hat{\mathcal{S}}(dx^1\otimes dx^1\otimes dx^3).
\end{equation}
Algebra of nontrivial symmetries is 1-dimensional:
\[
X=(x^3+x^1/2)\partial_2-(Bx^2+x^1/2)\partial_3.
\]
\end{enumerate}

\subsection{Metrics with $\tau(G)=5$ (5 types)}

From the technical viewpoint this case is the most complicated, since
it includes the largest number of cases under consideration.
This complexity is compensated by rareness  of the cases with
nontrivial symmetries.

\begin{enumerate}
\item $F=0,A_1=0,A_2=0,B_1=0,C_1=0.$ Canonical form of metric:
\begin{equation}\label{bm3s31}
G=dx^3\otimes dx^3\otimes dx^3+\hat{\mathcal{S}}(dx^1\otimes dx^3\otimes dx^3)+
\hat{\mathcal{S}}(dx^2\otimes dx^3\otimes dx^3)
+C_2\hat{\mathcal{S}}(dx^1\otimes dx^1\otimes dx^3)
\end{equation}
\[
+C_3\hat{\mathcal{S}}(dx^2\otimes dx^2\otimes dx^3).
\]
Algebra of nontrivial symmetries is 1-dimensional:
\[
X=(x^3+2C_3x^2)\partial_1-(x^3+2C_2x^1)\partial_2.
\]
\item$F=0,A_1=0,B_1=0,C_1=0,C_2=0.$ Canonical form of metric:
\begin{equation}\label{bm3s32}
G=dx^2\otimes dx^2\otimes dx^2+dx^3\otimes dx^3\otimes dx^3+\hat{\mathcal{S}}(dx^1\otimes dx^3\otimes dx^3)+
B_3\hat{\mathcal{S}}(dx^2\otimes dx^3\otimes dx^3)
\end{equation}
\[
+C_3\hat{\mathcal{S}}(dx^2\otimes dx^2\otimes dx^3).
\]
Algebra of nontrivial symmetries is 1-dimensional:
\[
X=((B_3С3-1)x^3+2(C_3^2-B_3)x^2-2x^1)\partial_1-C_3x^3\partial_2+x^3\partial_3.
\]
\item $A_1=0,A_2=0,A_3=0,B_1=0,C_1=0.$ Canonical form of metric:
\begin{equation}\label{bm3s33}
G=\hat{\mathcal{S}}(dx^1\otimes dx^2\otimes dx^3)+C_2\hat{\mathcal{S}}(dx^1\otimes dx^1\otimes
dx^3)+C_3\hat{\mathcal{S}}(dx^2\otimes dx^2\otimes dx^3)
+\hat{\mathcal{S}}(dx^1\otimes dx^3\otimes dx^3)
\end{equation}
\[
+\hat{\mathcal{S}}(dx^2\otimes dx^3\otimes dx^3).
\]
Algebra of nontrivial symmetries is 1-dimensional:
\[
X=(x^3+2x^1+2C_3x^2)\partial_1-(2x^2+2C_2x^1+x^3)\partial_2.
\]
\item$A_1=0,A_2=0,B_1=0,B_2=0,C_1=0.$ Canonical form of metric:
\begin{equation}\label{bm3s34}
G=\hat{\mathcal{S}}(dx^1\otimes dx^2\otimes dx^3)+C_2\hat{\mathcal{S}}(dx^1\otimes dx^1\otimes
dx^3)+C_3\hat{\mathcal{S}}(dx^2\otimes dx^2\otimes dx^3)
+\hat{\mathcal{S}}(dx^2\otimes dx^3\otimes dx^3)+
\end{equation}
\[
+dx^3\otimes dx^3\otimes dx^3.
\]
Algebra of nontrivial symmetries is 1-dimensional:
\[
X=(x^1+x^2/C_2)\partial_2-\frac{2x^1+2C_3x^2+x^3}{2C_2}\partial_1.
\]
\item$A_1=0,A_2=0,B_1=0,C_1=0,C_2=0.$ Canonical form of metric:
\begin{equation}\label{bm3s35}
G=\hat{\mathcal{S}}(dx^1\otimes dx^2\otimes dx^3)+dx^3\otimes dx^3\otimes
dx^3+\hat{\mathcal{S}}(dx^1\otimes dx^3\otimes dx^3)
+C_3\hat{\mathcal{S}}(dx^2\otimes dx^2\otimes dx^3)
\end{equation}
\[
+B_3\hat{\mathcal{S}}(dx^2\otimes dx^3\otimes dx^3).
\]
Algebra of nontrivial symmetries is 1-dimensional:
\[
X=(x^3+2x^2)\partial_2-(2x^1+2C_3x^2+B_3x^3)\partial_1.
\]
\end{enumerate}

\subsection{Metrics with $\tau(G)=6$ (1 type)}

There exists the only metric of general kind:
$A_1=0,A_2=0,B_1=0,C_1=0$:
\begin{equation}\label{bm3s36}
G=F\hat{\mathcal{S}}(dx^1\otimes dx^2\otimes dx^3)+\hat{\mathcal{S}}(dx^1\otimes dx^3\otimes dx^3)+\hat{\mathcal{S}}(dx^2\otimes dx^3\otimes
dx^3)+dx^3\otimes dx^3\otimes dx^3
\end{equation}
\[
C_2\hat{\mathcal{S}}(dx^1\otimes dx^1\otimes dx^3)+C_3\hat{\mathcal{S}}(dx^2\otimes dx^2\otimes
dx^3).
\]
with 1-dimensional algebra of nontrivial symmetries:
\[
X=(x^3+2Fx^1+2C_3x^2)\partial_1-(2Fx^2+2C_2x^1+x^3)\partial_2.
\]

\subsection{Metrics with $\tau(G)=7,8,9,10.$}\label{high}

Among the metrics of these affine types there is no metrics with
nontrivial symmetries.

So there are 41 general cubic homogeneous metrics of  different affine types, possessing nontrivial isometries.
Note, that our analysis deals only with general affine types.
Some general affine types with trivial isometries may contain special
metrics with some relations on its components, possessing
nontrivial isometries.
In majority cases such isometries will be equivalent to one of the
isometries, associated  with  considered metrics with $\tau(G)\le 6.$
In these cases we have equivalent metrics.
However, it is possible the situation when under some particular values
of metric components the metric will not be equivalent to any of
above considered ones.
Such "very special"\, metrics come out from the scope of our
investigation (see, however, the table at the end of the section \ref{projj}).

\section{Affine-invariant classification}

Some of the affine types with nontrivial symmetries are, in fact,
affine-equivalent.
In order to clear the question on equivalence of the above listed 41 classes, let us
turn to (affine-)invariant properties of their symmetries fields.
Preliminary  classifying can be  carried out by dimension of symmetries algebra.
Combining different affine types possessing equal dimensions of symmetry algebra,
we go to the following  non-equivalent classes:

\begin{enumerate}
\item
class of affine types with 2-dimensional algebra of symmetry,
including the cases (first number  is affine type, second number is  order number in correspondent section):
1.1, 2.2, 2.5, 2.6, 3.3, 3.5, 3.9, 3.13, 4.9;
\item
class of affine types with 1-dimensional algebra of symmetry,
including the cases: 2.1, 2.3, 2.4, 3.1, 3.2, 3.4, 3.6, 3.7, 3.8, 3.10, 3.11, 3.12,
4.1,4.2, 4.3, 4.4, 4.5, 4.6, 4.7, 4.8, 4.10, 5.1, 5.2, 5.3, 5.4, 5.5, 6.1.
\item
types 1.2, 1.3, 2.7, 2.8, 2.9 with infinitely-dimensional algebra of symmetries;
\item
all types, without nontrivial symmetries;
\item
"very special metrics", which have not been included in previous items.
\end{enumerate}

The two last classes come out of the scope of our investigation.
The first two classes admit further more detailed classifying.
Direct calculation shows, that commutators of the pair of symmetry
fields for metrics of the first class are:

\begin{enumerate}
\item 0, for the cases 1.1, 2.2, 2.5, 3.3, 3.5;
\item
$(3/2)X_2$ for the cases 2.6, 3.9, 3.13, 4.9.
\end{enumerate}

So, we conclude,  that {\it groups of metrics $\{1.1,2.2,2.5,3.3,3.5\},$ and
$\{2.6, 3.9, 3.13, 4.9\}$ are affine-nonequivalent.}
The question on affine equivalency of metrics inside these groups
remaines opened. We come back to this question in next section.

Let us go to the affine types with 1-dimensional algebras of
symmetry.
Rough classifying of these types may be carried out by comparing of
the simplest affine invariant of  their algebras --- divergencies of
corresponding vector fields: $\text{div}\, X\equiv\partial_iX^i.$
Elementary calculations show, that $\text{div}\, X=0$ for the following cases:
2.1, 2.5, 3.1, 3.2, 3.4, 3.6, 3.8,  3.11, 3.12, 4.1, 4.2, 4.3, 4.4, 4.7, 4.10,  5.1, 5.3, 5.4, 5.5, 6.1
and  $\text{div}\,X=\text{const}\neq0$ for the cases 2.3, 2.4, 3.7, 3.10, 4.5, 4.6, 4.8, 5.2.
So, {\it affine types,  lying in these different groups,  are affine-nonequivalent.}

Further more detailed classifying of the metrics inside these groups
implies comparing of other affine invariants.
Since all considered symmetries are described by linear vector
fields, let us consider the following {\it matrix of vector field $A$}, defined by relation:
\[
X^\alpha=A^\alpha_\beta x^\beta,
\]
where $A_\beta^\alpha$ --- components of $A.$ This definition means, that
$A$ is affine tensor of valency  $(1,1).$ Its affine invariants are the following quantities:
\[
I_1\equiv\text{Tr}(A),\quad\dots, I_n\equiv\text{Tr}(A^n);\quad \Delta\equiv\det(A).
\]
Note, that $\text{div}\,X=I_1.$ Equivalent metrics must satisfy  {\it colinearity conditions}:
\begin{equation}\label{coll}
\sqrt[n]{\frac{I_n}{I'_n}}=\sqrt[3]{\frac{\Delta}{\Delta'}}=C=\text{const}
\end{equation}
for all  $n=1,\dots,$
where $\{I_n,\Delta\}$ is the system of invariants of one metric, $\{I'_n,\Delta'\}$ is
the system of invariants for another one.
It is possible to construct other invariants, but the set $\{I_n\}$
is sufficient for our purposes.

For the metrics with $\text{div} X\neq0$ matrices of their vector fields and invariants have the following kind:
\[
1.2:\quad
\left(
\begin{array}{ccc}
-2&0&0\\
0&1&0\\
0&0&0
\end{array}
\right),\  I_n=1+(-2)^n;\quad
2.3:\quad
\left(
\begin{array}{ccc}
0&0&0\\
0&1&0\\
0&0&-1/2
\end{array}
\right),\  I_n=\frac{1+(-2)^n}{(-2)^n};
\]
\[
2.4:\quad
\left(
\begin{array}{ccc}
1&0&0\\
-1&-2&0\\
0&0&0
\end{array}
\right),\  I_n=1+(-2)^n;\quad
3.7:\quad
\left(
\begin{array}{ccc}
1&0&0\\
0&0&0\\
-1&0&-2
\end{array}
\right),\  I_n=1+(-2)^n;
\]
\[
3.10:\quad
\left(
\begin{array}{ccc}
0&0&0\\
0&1&0\\
-2\epsilon &0&-2
\end{array}
\right),\  I_n=1+(-2)^n;\quad
4.5:\quad
\left(
\begin{array}{ccc}
1&0&0\\
-B&0&0\\
0&2B^2&-2
\end{array}
\right),\  I_n=1+(-2)^n;
\]
\[
4.6:\quad
\left(
\begin{array}{ccc}
0&0&0\\
0&1&0\\
2B&1&-2
\end{array}
\right),\  I_n=1+(-2)^n;\quad
2.3:\quad
\left(
\begin{array}{ccc}
0&0&-C\\
2(C^2\mp1)&-2&\pm C\\
0&0&1
\end{array}
\right),\  I_n=1+(-2)^n;
\]
\[
5.2:\quad
\left(
\begin{array}{ccc}
-2&2(C_3^2-B_3)&B_3C_3-1\\
0&0&-C_3\\
0&0&1
\end{array}
\right),\  I_n=1+(-2)^n.
\]
Obviously, that conditions (\ref{coll})
are satisfied for all metrics from the group with $\text{div}\, X=I_1\neq0.$

For the group with  $\text{div}\, X=I_1=0$ matrices of their vector fields and invariants have the following
kind:
\[
2.1:\quad
\left(
\begin{array}{ccc}
0&0&0\\
0&1&0\\
0&0&-1
\end{array}
\right),\  I_n=1+(-1)^n;\quad
3.1:\quad
\left(
\begin{array}{ccc}
0&0&0\\
0&1&0\\
0&-1&-1
\end{array}
\right),\  I_n=1+(-1)^n;\quad
\]
\[
3.2:\quad
\left(
\begin{array}{ccc}
0&0&0\\
0&1&0\\
-1/2&0&-1
\end{array}
\right),\  I_n=1+(-1)^n;
3.4:\quad
\left(
\begin{array}{ccc}
1&0&1\\
0&0&0\\
0&-1/2&-1
\end{array}
\right),\  I_n=1+(-1)^n;\]
\[
3.6:\quad
\left(
\begin{array}{ccc}
-1&-1/2&0\\
0&0&0\\
0&1/2&1
\end{array}
\right),\  I_n=1+(-1)^n;
\]
\[
3.8:\quad
\left(
\begin{array}{ccc}
0&0&0\\
0&0&1\\
0&-\epsilon_1\epsilon_2&0
\end{array}
\right),\  I_n=(-\epsilon_1\epsilon_2)^{n/2}(1+(-1)^n);
\]
\[
3.11:\quad
\left(
\begin{array}{ccc}
0&0&0\\
0&0&1\\
\mp1/2&\mp1&0
\end{array}
\right),\  I_n=(\mp1)^{n/2}(1+(-1)^n);
\]
\[
3.12:\quad
\left(
\begin{array}{ccc}
0&0&1\\
0&0&0\\
-1&-1/2&0
\end{array}
\right),\  I_n=(-1)^{n/2}(1+(-1)^n);\quad
\]
\[
4.1:\quad
\left(
\begin{array}{ccc}
0&0&0\\
0&\epsilon_2F&1\\
0&-\epsilon_1\epsilon_2&-\epsilon_2F
\end{array}
\right),\  I_n=(F^2-\epsilon_1\epsilon_2)^{n/2}(1+(-1)^n);
\]
\[
4.2:\quad
\left(
\begin{array}{ccc}
0&0&0\\
-1/2F&-1&0\\
0&\pm1/F&1
\end{array}
\right),\  I_n=(1+(-1)^n);\quad
\]
\[
4.3:\quad
\left(
\begin{array}{ccc}
0&0&0\\
\mp1/2&\mp F&\mp1\\
0&1&\pm F
\end{array}
\right),\  I_n=(F^2\mp1)^{n/2}(1+(-1)^n);
\]
\[
4.4:\quad
\left(
\begin{array}{ccc}
1&0&1/2F\\
-1/F&-1&-1/2F\\
0&0&0
\end{array}
\right),\  I_n=1+(-1)^n;\quad
\]
\[
4.7:\quad
\left(
\begin{array}{ccc}
0&0&0\\
0&0&1\\
-\epsilon_2C/2&-\epsilon_1\epsilon_2&0
\end{array}
\right),\  I_n=(-\epsilon_1\epsilon_2)^{n/2}(1+(-1)^n);
\]
\[
4.10:\quad
\left(
\begin{array}{ccc}
0&0&0\\
1/2&0&1\\
-1/2&-B&0
\end{array}
\right),\  I_n=(-B)^{n/2}(1+(-1)^n);
\]
\[
5.1\quad
\left(
\begin{array}{ccc}
0&2C_3&1\\
-2C_2&0&-1\\
0&0&0
\end{array}
\right),\  I_n=(-4C_2C_3)^{n/2}(1+(-1)^n);\quad
\]
\[
5.3:\quad
\left(
\begin{array}{ccc}
2&2C_3&1\\
-2C_2&-2&-1\\
0&0&0
\end{array}
\right),\  I_n=(4(1-C_2C_3))^{n/2}((-1)^n+1);
\]
\[
5.4\quad
\left(
\begin{array}{ccc}
-1/C_2&-C_3/C_2&-1/2C_2\\
1&1/C_2&0\\
0&0&0
\end{array}
\right),\  I_n=(1-C_2C_3)^{n/2}(1+(-1)^n);\quad
\]
\[
5.5:\quad
\left(
\begin{array}{ccc}
-2&-2C_3&-B_3\\
0&2&1\\
0&0&0
\end{array}
\right),\  I_n=4^{n/2}(1+(-1)^n);
\]
\[
6.1\quad
\left(
\begin{array}{ccc}
2F&2C_3&-B_3\\
-2C_2&-2F&-1\\
0&0&0
\end{array}
\right),\  I_n=2^n(F^2-C_2C_3)^{n/2}(1+(-1)^n)
\]
Comparison of the series of invariants leads to the  following
potential classes of affine equivalent metrics:

\begin{enumerate}
\item $\{2.1, 3.1, 3.2, 3.4, 3.6, 3.8 (\epsilon_1\epsilon_2<0)$, 3.11 ($-$ in the metric), 4.1 $(F^2>\epsilon_1\epsilon_2)$,
4.2, 4.3 $(F^2>\pm1)$, 4.4, 4.7 $(\epsilon_1\epsilon_2<0)$, 4.10 $(B<0)$, 5.1 $(C_2C_3<0)$, 5.3\, $C_2C_3<1,$ 5.4
$(C_2C_3<1)$, 5.5, 6.1 $(F^2>C_2C_3)$\};
\item
\{3.8 $(\epsilon_1\epsilon_2>0)$, 3.11 ($+$ in the metric), 3.12, 4.1 $(F^2<\epsilon_1\epsilon_2)$,
 4.3 $(F^2<\pm1)$, 4.7 $(\epsilon_1\epsilon_2>0)$, 4.10 $(B<0)$, 5.1 $(C_2C_3>0)$, 5.3\, $C_2C_3>1,$ 5.4
$(C_2C_3<1)$, 6.1 $(F^2<C_2C_3)$\};
\item
\{4.1 $(F=\epsilon_1\epsilon_2)$,
 4.3 $(F^2=+1)$ ("+"\, in the metric), 4.10 $B=0$, 5.3\, $C_2C_3=1,$,  5.4
$(C_2C_3=1)$, 6.1 $(F^2=C_2C_3)$\};
\end{enumerate}

The more detailed additional investigation of the cases 3, when all
invariants formally vanish, gives the following corrections:

\begin{enumerate}
\item The metric 4.1 under $F^2=\epsilon_1\epsilon_2$ admits symmetry vector field with one
arbitrary function of all coordinates, i.e.
admits infinitely-dimensional group of symmetry.
\item
The metrics  4.3 under $F^2=1,$  4.10 under $B=0$, 5.3\,,  5.4 under $C_2C_3=1$ and 6.1 under $F^2=C_2C_3$ admit
2-dimensional nonabelian group of symmetries with commutator of the
kind:
$[X_1,X_2]=(3/2)X_2.$
\end{enumerate}

Resuming our investigation, we conclude, that
{\it all homogeneous cubic metrics of general affine types
are divided on  8 affine-nonequivalent classes:}
\begin{enumerate}
\item
class $\{1.1, 2.2, 2.5, 3.3, 3.5\}$ (2-dimensional abelian algebra of nontrivial symmetries);
\item
class $\{2.6, 3.9, 3.13, 4.9, 4.3\, (F^2=1),  4.10\,  (B=0), 5.3,5.4\,  (C_2C_3=1), 6.1\,  (F^2=C_2C_3)\}$
(2-dimensional nonabelian algebra of nontrivial symmetries);
\item
class $\{1.2,1.3, 2.7, 2.8, 2.9, 4.1\, (F^2=\epsilon_1\epsilon_2)\}$ (infinitely-dimensional algebra of isometries);
one can subdivide this class on the following subclasses:  (1):  $\infty^2$-dimensional  group (1.3),
(2): $\infty$-dimensional group (4.1,2.8,2.9) and (3): $\infty+1$-dimensional group (1.2, 2.7);
\item
class  $\{2.3,2.4,3.7,3.10,4.5,4.6,4.8,5.2\}$ (1-dimensional algebra of nontrivial symmetries with $I_1\neq0$);
\item
class \{2.1, 3.1, 3.2, 3.4, 3.6, 3.8 $(\epsilon_1\epsilon_2<0)$, 3.11 ($-$ in metric), 4.1 $(F^2>\epsilon_1\epsilon_2)$,
4.2, 4.3 $(F^2>\pm1)$, 4.4, 4.7 $(\epsilon_1\epsilon_2<0)$, 4.10 $(B<0)$, 5.1 $(C_2C_3<0)$, 5.4
$(C_2C_3<1)$, 5.5, 6.1 $(F^2>C_2C_3)$\} (1-dimensional algebra of nontrivial symmetries,
$I_n=C^{n/2}(1+(-1)^n),\ C=\text{const}>0$;
\item
class \{3.8 $(\epsilon_1\epsilon_2>0)$, 3.11 ($+$ in metric), 3.12, 4.1 $(F^2<\epsilon_1\epsilon_2)$,
 4.3 $(F^2<\pm1)$, 4.7 $(\epsilon_1\epsilon_2>0)$, 4.10 $(B<0)$, 5.1 $(C_2C_3>0)$, 5.4
$(C_2C_3<1)$, 6.1 $(F^2<C_2C_3)$\}; (1-dimensional algebra of nontrivial symmetries,
$I_n=(-C)^{n/2}(1+(-1)^n),$ $C=\text{const}<0$);
\item
class of metrics with nontrivial symmetries, which are absent in previous list;
\item class of metrics without symmetries.
\end{enumerate}

The question on affine equivalence of the metrics inside these
classes is opened. Next section we'll prove that the answer is,
generally speaking,
negative.

\section{Connection with projective classification}\label{projj}

Lets clear connection of obtained results with well known projective
classification of cubic 3-dimensional forms \cite{sokol}.
Combination of methods of projective geometry and cubic matrix
algebra leads to the following classifying theorem.

\bigskip
{\bf THEOREM (on classification of real cubic forms)} {\sf Any cubic form over field of real numbers
belong to one of the classes of real affine-equivalency (only nonzero components of canonical kind of cubic metric are presented):
\begin{enumerate}
\item general class $A_1=A_2=A_3=1,$
with 10 nonequivalent subclasses: $F<-(\sqrt3+1)/2,$ $F=-(\sqrt3+1)/2$,
$-(\sqrt3+1)/2<F<-1/2,$ $-1/2<F<0,$ $F=0,$ $0<F<(\sqrt3-1)/2,$ $F=(\sqrt3-1)/2,$
$(\sqrt3-1)/2<F<1,$ $F=1,$ $F>1.$
\item Degenerated class  I: $A_1=A_2=F=1;$
\item
Degenerated class  II: $A_1=F=1;$
\item
Degenerated class  III: $F=1;$
\item
Degenerated class IV: $A_1=C_3=1;$
\item
Degenerated class  V: $C_1=C_3=1;$
\item
Degenerated class  VI: $A_1=A_2=1;$
\item
Degenerated class  VII: $C_1=1;$
\item
Degenerated class  VIII: $A_1=1;$
\item
Degenerated class  IX: $A_3=C_1=B_3=1;$
\item
Degenerated class X: $-A_2=C_1=B_3=1;$
\item
Degenerated class XI: $A_2=C_1=B_3=1;$
\item
Degenerated class XII: $C_1=B_3=1;$
\item
Degenerated class XI: $-A_2=C_1=1.$
\end{enumerate}
}

Comparison of these canonical types with classes of isometries leads
to the following conclusions:
\begin{enumerate}
\item General class under $F\neq -1/2$ has no nontrivial symmetries and so it belongs to symmetry class 8.
In case $F=-1/2$ generic metric acquires  2-dimensional  abelian group
of symmetries and can be related to the symmetry class 1;
\item Degenerated class I has no nontrivial symmetries and so it belongs to symmetry class
8;
\item Degenerated class II has 1-dimensional group with
$I_1=0$ and is related to symmetry class  5;
\item
Degenerated class III has 2-dimensional abelian group and is
related to symmetry class 1;
\item
Degenerated class IV has 1-dimensional group with
$I_1\neq0$ and is related to symmetry class  4;
\item
Degenerated class V has 2-dimensional nonabelian group with
is related to symmetry class  2;
\item
Degenerated class VI has $\infty$-dimensional group
and is related to symmetry class  3(2);
\item
Degenerated class VII has $\infty+1$-dimensional group
and is related to symmetry class  3(3)
\item
Degenerated class VIII has $\infty$-dimensional group
and is related to symmetry class  3(1);
\item
Degenerated class IX has no nontrivial symmetries and is related to symmetry class
9;
\item
Degenerated class X has 1-dimensional group with
$I_1=0$ and is related to symmetry class  5;
\item
Degenerated class XI has 1-dimensional group with
$I_1=0$ and is related to symmetry class  5;
\item
Degenerated class XII has 2-dimensional abelian group and is
related to symmetry class 1;
\item
Degenerated class XIII has $\infty$-dimensional group
and is related to symmetry class  3(2).
\end{enumerate}

Interrelations between symmetry and projective classifications are
resumed in the following table.
\bigskip

\hspace{0.5em}{\small
\begin{tabular}{|c|c|c|c|c|c|c|c|c|}
\hline
Symmetries classes& 1&2&3&4&5&6&7&8\\
\hline
Projective classes&III,XII&V&(1): VIII, (2): & IV&
II,X,XI&?&---&Gen, I,IX\\
&&&VI,XIII, (3): VII&&&&&\\
\hline
\end{tabular}
}

\medskip
Analysis of the table leads to the following important conclusions:

\begin{enumerate}
\item Symmetries classification is more rough, then projective, since some classes of
symmetries contain several non-equivalent projective classes.
\item Emptiness of the column with number 7 means, that we have studied in fact  all non-equivalent classes of cubic
metrics.
\item Emptiness of the column with number 6 means that  5-th and 6-th symmetries classes are identical.
Common constant in righthand side of the colinearity
condition (\ref{coll})
between  these classes will be imaginary. This corresponds to the statement, that isometries fields form not $R$-module, as we have
assumed,
but  $C$-module.
\end{enumerate}

\bigskip

Author is grateful to D.G. Pavlov for stimulating discussion and
financial supporting of this work.

\end{document}